\documentstyle[prl,aps]{revtex}  
\def\bra{\langle} \def\ket{\rangle} 
\begin{document} 
\draft 

\title{Grover Algorithm with  zero theoretical failure rate} 
\author{G. L. Long$^{1,2,3,4}$}
\address{$^1$Department of Physics, Tsinghua University, Beijing 100084, P.R.China\\
$^2$Key Laboratory for Quantum Information and Measurements, Ministry of Education, 
Beijing 100084,  P.R. China\\
$^3$Institute of Theoretical Physics, Chinese Academy of Sciences, Beijing, 100080, P. R. 
China\\
$^4$Center of Atomic, Molecular and Nanosciences, Tsinghua University, Beijing 100084, P. 
R. China}
\date{\today}
\maketitle 
 
\date{\today} 
\maketitle 
 
\begin{abstract} 
In standard Grover's algorithm for quantum searching,  the probability of finding the 
marked item is not exactly 1. In this Letter we present a modified version of Grover's 
algorithm that searches a marked state with full successful rate. The modification is done 
by replacing the  phase inversion by two phase rotation through angle $\phi$. The rotation 
angle is given analytically to be $\phi=2 \arcsin\left(\sin{\pi\over (4J+6)}\over 
\sin\beta\right)$, where $\sin\beta={1\over \sqrt{N}}$, $N$  the number of items in the 
database, and $J$ an integer equal to or greater than the integer part of $({\pi\over 
2}-\beta)/(2\beta)$. Upon measurement at $(J+1)$-th iteration, the marked state is 
obtained with certainty.
\end{abstract} 
 

Grover's quantum search algorithm\cite{r1} is an important development in quantum 
computation.  It achieves square-root speedup over classical algorithms in unsorted 
database searching. It has extensive applications, because  many problems, for instance 
deciphering the DES encryption scheme, can be reduced to this problem\cite{science}. 
Starting from an evenly distributed state, Grover algorithm searches the database with
\begin{eqnarray}
j_{op}=[(\pi/2-\beta)/(2\beta)],
\label{jop}
\end{eqnarray}
or $j_{op}+1$ number of times,  whichever of $(2(j_{op}+1)+1)\beta$ and $(2j_{op}+1)\beta$ 
is closest to ${\pi\over 2}$. Here $\beta=\arcsin{1\over \sqrt{N}}$ and $[]$ means taking 
the integer part. $N$ is the number of items in the database. Maximum probability  is 
achieved when measurement is made at the optimal iteration step, and it is
\begin{eqnarray}
P_{max}=\sin^2((2j_{op}+1)\beta)\cong 1.
\end{eqnarray}
It equals to 1 if $(2j_{op}+1)\beta={\pi\over 2}$. This condition is usually approximately 
satisfied. This can be seen from table \ref{t1} where values of $j_{op}$, 
$(2j_{op}+1)\beta$ for $N$ are given.  The deviation  $(2j_{op}+1)\beta$ from ${\pi\over 
2}$ is in the order of ${1\over \sqrt{N}}$. This small deviation becomes negligible when 
the dimension of the quantum database becomes very large, for instance in deciphering the 
DES code where $N=2^{56}$ the deviation is only $3\times 10^{-9}$. Standard Grover 
algorithm has already achieved high probability, and in most potential applications it is 
sufficient. 

However, in problems where certainty is vital, especially when the dimension is not so 
big, using a searching algorithm with certainty becomes important. Meanwhile constructing 
such a quantum search algorithm itself is an interesting issue. In this Letter we present 
such a modified Grover algorithm. 

In fact, two such algorithms already exist. One is given by Brassard et al\cite{r3} where 
the generalized algorithm searches the database $j_{op}$ iterations with the standard 
Grover algorithm, and then run one more iteration with a modified algorithm whose step is 
smaller. H{\o}yer gave another generalization\cite{r4}, where certainty is achieved by 
modifying the Grover algorithm and making change to the initial distribution.  Our 
algorithm here complements with these algorithms. In addition, the present algorithm 
materializes one earlier anticipation\cite{r5}.  It was pointed out that when the phase 
inversions are replaced by arbitrary rotations in Grover's algorithm\cite{r6}, a quantum 
search algorithm with a smaller iteration can be constructed. Zalka anticipated that  this 
could be used to achive certainty in quantum searching\cite{r5} by running a quantum 
searching algorithm with a smaller step so that at an integer number of iteration, the 
quantum computer state vector is exactly the marked state. Our algorithm here is just such 
an algorithm.

The generalized Grover algorithm here starts from the evenly distributed state
\begin{eqnarray}
|\psi_i\ket={1\over \sqrt{N}}\sum_{i}|j\ket={1\over \sqrt{N}}(|0\ket +|1\ket+...+|\tau\ket 
+...+|N-1\ket)=\sin\beta|1\ket+\cos\beta|2\ket,
\label{ini}
\end{eqnarray}
and the searching operator is
\begin{eqnarray}
Q=-W\;I_0\;W\;I_{\tau}=\left[\begin{array}{cc}
-e^{i\phi}(1+(e^{i\phi}-1)\sin^2\beta)  &-(e^{i\phi}-1)\sin\beta\cos\beta \\
 -e^{i\phi} (e^{i \phi}-1)\sin\beta\cos\beta& 
-e^{i\phi}+(e^{i\phi}-1)\sin^2\beta\end{array}
\right],
\label{q}
\end{eqnarray}
where the matrix expression is written in the following basis
\begin{eqnarray}
|1\ket&=&|\tau\ket,\nonumber\\
|2\ket&=&{1\over \sqrt{N-1}}\sum_{i\ne \tau}|i\ket,
\label{basis}
\end{eqnarray}
and
\begin{eqnarray}
I_{\tau}&=&I+(e^{i\phi}-1)|\tau\ket\bra \tau|,\nonumber\\
I_{0}&=&I+(e^{i\phi}-1)|0\ket\bra 0| ,
\label{tau}
\end{eqnarray} 
where
\begin{eqnarray}
\phi&=& 2 \arcsin\left( \sin\left(\pi \over 4J+6\right) \over \sin\beta\right)\nonumber,\\
J&\ge&J_{op}.
\label{phase}
\end{eqnarray}
Here the two phase rotations are equal which is required by the phase matching 
condition\cite{r7,r7p}. Certainty in quantum searching is achieved by measuring the 
quantum computer at $J+1$ iteration. In standard Grover algorithm, $\phi=\pi$.  In table 
\ref{t2}, we gave the angle $\phi$ for some values of $N$. It is seen that in general the 
phase rotations are very close to $\pi$. We see that at small $N$, the deviation of $\phi$ 
to $\pi$ is big, and it decreases when $N$ becomes large. The certainty of the algorithm 
can be examined by direct computation. We will give the detailed derivation of this result 
in the $SO(3)$ picture of quantum searching algorithm\cite{r8}. 

Equation (\ref{phase}) has a real solutions for $J\geq j_{op}$, otherwise the solution 
will be complex. An integer $J\geq j_{op}$ fixes an  phase rotation that searches the 
marked state with certainty in $J+1$ steps. The lower bound $j_{op}$ tells us that it can 
not be faster than the standard Grover algorithm. $J$ can be chosen to be $j_{op}$, or an 
integer larger than $j_{op}$ for convenience.

In the following part, we show the above result and give the expression for the 
probability  during the searching process.  We do this in the $SO(3)$ picture introduced 
recently\cite{r8}. In this picture, the quantum search operator (\ref{q}) corresponds to a 
rotation in space 
\begin{eqnarray}
R_Q=\left[\begin{array}{ccc}R_{11} & R_{12} & R_{13}\\
                            R_{21} & R_{22} & R_{23}\\
							R_{31} & R_{32} & R_{33}\end{array}\right],
\end{eqnarray}
where 
\begin{eqnarray}
R_{11}&=&\cos\phi (\cos^22\beta\cos\phi+\sin^2 2\beta)+\cos2\beta\sin^2\phi),\nonumber\\
R_{12}&=&\cos\phi\sin\phi (\cos2\beta-1),\nonumber\\
R_{13}&=&-\cos\phi \sin4\beta \sin^2{\phi \over 2}+\sin 2\beta\sin^2\phi,\nonumber\\
R_{21}&=&-\cos2\beta\cos\phi\sin\phi+\left(cos^2{\phi\over 2}-\cos4\beta\sin^2{\phi\over 
2}\right)\sin\phi,\nonumber\\
R_{22}&=&\cos^\phi+\cos2\beta \sin^2\phi,\nonumber\\
R_{23}&=&-\cos\phi\sin2\beta\sin\phi-\sin4\beta\sin^2{\phi\over 2}\sin\phi,\nonumber\\
R_{31}&=&-\sin4\beta\sin^2{\theta \over 2},\nonumber\\
R_{32}&=&\sin2\beta\sin\phi,\nonumber\\
R_{33}&=&\cos^22\beta+\cos\phi\sin^22\beta.\nonumber
\end{eqnarray}
The above rotation is a rotation about the following axis
\begin{eqnarray}
\vec{l}=\left(\begin{array}{c}\cos{\phi\over 2}\\ \sin{\phi\over 2}\\ \cos{\phi\over 
2}\tan{\beta}\end{array}\right),
\label{axis}
\end{eqnarray}
through an angle $\alpha$
\begin{eqnarray}
\alpha=4\arcsin(\sin({\phi\over 2})\sin\beta).
\end{eqnarray}
State vector $|\psi\ket=(a+b i)|1\ket+(c+d i)|2\ket$ is represented by the polarization 
vector
\begin{eqnarray}
\vec{r}_{\psi}=\bra \psi|\vec{\sigma}|\psi\ket=\left(\begin{array}{c}
2(ac+bd)\\ 2(-bc+ad) \\ a^2+b^2-c^2-d^2\end{array}\right),
\label{polar}
\end{eqnarray}
where $\vec{\sigma}=\sigma_x \vec{i}+\sigma_y\vec{j}+\sigma_z\vec{k}$, and $\vec{i}$, 
$\vec{j}$ and $\vec{k}$ are the unit vectors along the $x$-, $y$-, and $z$-axis.
The initial state $|\psi_i\ket$ and the marked state $|\tau\ket$ are represented by
\begin{eqnarray}
\vec{r}_i=\left(\begin{array}{c} \sin(2\beta)\\ 0\\ -\cos(2\beta)\end{array}\right),
\vec{r}_f=\left(\begin{array}{c} 0 \\ 0\\ 1\end{array}\right).
\end{eqnarray}
Now we want to find out the angle that we must rotate to shift $\vec{r}_i$ to $\vec{r}_f$.
The equation for a line passing through the origin and parallel to the rotational axis is
\begin{eqnarray}
{x\over \cos{\phi \over 2}}={y\over \sin{\phi\over 2}}={z\over \cos{\phi\over 
2}\tan\beta},
\label{line}
\end{eqnarray}
and the equation for the plane passing through $(0,0,1)^T$ and normal to the rotational 
axis is
\begin{eqnarray}
x\cos{\phi\over 2}+y\sin{\phi\over 2}+(z-1)\cos{\phi\over 2}\tan\beta=0.
\label{plane}
\end{eqnarray}
The intersecting point of (\ref{line}) with (\ref{plane}) is
\begin{eqnarray}
\vec{r}_o=\left(\begin{array}{c}
c\cos^2({\phi\over 2})\tan\beta\\
c\sin({\phi\over 2})\cos({\phi\over 2})\tan\beta\\
c\cos^2({\phi\over 2})\tan^2\beta\end{array}\right),
\label{point}
\end{eqnarray}
where $c={1\over 1+\cos^2{\phi\over 2}\tan^2\beta}$.
The angle $\omega$ between $\vec{r}_i-\vec{r}_0$ and $\vec{r}_f-\vec{r}_0$ is the angle we 
have to rotate in a given number of iterations. Using
\begin{eqnarray}
(\vec{r}_i-\vec{r}_o)\cdot(\vec{r}_f-\vec{r}_o)=|\vec{r}_i-\vec{r}_0||\vec{r}_f-\vec{r}_0|
\cos\omega,
\end{eqnarray}
we find that
\begin{eqnarray}
\cos\omega=-\cos^2\beta-\cos\phi\sin^2\beta=\cos(2\arccos x),
\end{eqnarray}
where
\begin{eqnarray}
x=\sin({\phi\over 2})\sin\beta.
\end{eqnarray}
Certainty in finding the marked state is achieved if angle $\omega$ is $J+1$ times of the 
basic rotation angle $\alpha$:
\begin{eqnarray}
\omega=2\arccos(x)=(J+1) \alpha=4 (J+1)\arcsin(x).
\label{equation}
\end{eqnarray}
 Using the trigonometric relation $\arcsin x+\arccos x={\pi\over 2}$, we obtain 
\begin{displaymath}
\omega=2 ({\pi\over 2}-\arcsin x)=4(J+1)\arcsin(x).
\end{displaymath}
This gives the result of equation (\ref{phase}). Equation (\ref{phase}) has  real 
solutions for $J \geq j_{op}$. $J=j_{op}$ is the minimum in most cases. In the cases of 
$N=4$ and $N=1$,  $J=j_{op}-1$ itself is  a solution.

The probability for finding the marked state during the searching can be obtained easily. 
In the $SO(3)$ picture, the polarization vector at a given iteration is obtained by a 
simple geometric argument, 
\begin{eqnarray}
\vec{r}_j=\vec{r}_i\cos\omega+\vec{l}_n(\vec{l}_n\cdot\vec{r}_i)(1-
\cos\omega)+(\vec{l}_n\otimes \vec{r}_i)\sin\omega,
\label{rstate}
\end{eqnarray}
where $\vec{l}_n$ is the rotational axis (\ref{axis}) normalized to unity.
Using the equation (\ref{polar}), the state vector can be determined easily. The 
probability for finding the marked state is $(z+1)/2$. 

We can also write out the expressions in the $U(2)$ formalism. After diagonalization, the 
$Q$ operator can be written as,
\begin{eqnarray}
Q&=&T\;\Lambda\;T^{\dagger},\nonumber
\end{eqnarray}
where
\begin{eqnarray}
T&=& {1\over \sqrt{N_T}}\left(\begin{array}{cc} e^{-i {\phi \over 2}}(\cos({\phi\over 
2})\sin\beta+\cos\beta') & -\cos\beta\\
  \cos\beta & e^{i {\phi \over 2}}(\cos({\phi\over 2})\sin\beta+\cos\beta')   
\end{array}\right),\nonumber\\
\Lambda&=&\left(\begin{array}{cc}-e^{i (\phi+2\beta')} & 0\\
                         0                & -e^{i (\phi-2\beta')}\end{array}
\right),\nonumber\\
\beta'&=&\alpha/4=\arcsin(\sin({\phi\over 2})\sin\beta),\nonumber\\
N_T&=&\cos^2\beta+(\cos({\phi\over 2})\sin\beta+\cos\beta')^2.\nonumber
\end{eqnarray}
Successive operations of $Q$ can be written analytically through
\begin{displaymath}
Q^n=T\;\Lambda^n\;T^{\dagger}.
\end{displaymath}

In summary, a Grover algorithm with certainty is present.  Together with  the algorithms 
in Ref.\cite{r3} and Ref.\cite{r4}, there are 3 choices of quantum searching algorithm for  
finding the marked state with certainty. Our algorithm may be appreciated in cases where 
the dimension is not big and certainty is important, and in cases where  preparation of 
initial state and the change of the experimental setting during the computation process 
are difficult.

This work is supported by the Major State Basic Research Developed Program 
Grant No. G200077400, the China National Natural Science Foundation Grant 
No. 60073009, the Fok Ying Tung Education Foundation, and the 
Excellent Young University Teachers' Fund of Education Ministry of China.

\begin{table} 
\begin{center}
\caption{Examples of $j_{op}$ and $(2j_{op}+1)\beta$}
\label{t1}
\begin{tabular}{ccccccccccc}\\ 
$N$ & 2 & 4 & 8 &  100 & 1000 & $10^4$ & $10^6$ & $10^8$ & $10^{10}$& $2^{56}$\\
$j_{op}$ &0 &1 & 1 & 7 & 24 & 78 & 784 & 7853 & 78539& 210828713\\
${(2j_{op}+1)\beta \over {\pi\over 2}}$ & ${1\over 2}$ & 1 & 0.69016  &0.956528 &0.986617 
&0.99951 &0.998857 &0.999939 & 0.999996 & 0.999999997\\
\end{tabular}
\end{center}
\end{table}

\begin{table} 
\begin{center}
\caption{Examples of $j_{op}+1$ and $\phi$}
\label{t2}
\begin{tabular}{ccccccccccc}\\ 
$N$ & 2 & 4 & 8 & 16 & 100 & 1000 & $10^4$ & $10^6$ & $10^8$ & $10^{10}$\\
$j_{op}$+1 & 1 &1 & 2 & 3 & 8 & 25 & 79 & 785 & 7854 & 78540\\
${\phi \over \pi}$ & ${1\over 2}$ & 1 & 0.677007 & 0.698709 & 0.748018 & 0.854022& 0.90089 
&0.989752 & 0.992688 & 0.9973\\
\end{tabular}
\end{center}
\end{table}

\begin{thebibliography}{99}
\bibitem{r7p} Note that the phase condition in Ref.\cite{r4} is different from this one, 
because his initial state is different.
\end{thebibliography}
\end{document}